\documentclass[10pt,a4paper]{article}
\usepackage{latexsym}
\usepackage{fullpage}
\usepackage{times}
\usepackage{natbib}
\usepackage[dvips]{epsfig}
\usepackage[dvips]{graphicx}
\usepackage[english]{babel}
\usepackage{amssymb,amsmath,array}
\usepackage{multirow}
\usepackage{color}

\usepackage{setspace}
\makeatletter
\def\@normalsize{\@setsize\normalsize{12pt}\xpt\@xpt
\abovedisplayskip 10pt plus2pt minus5pt\belowdisplayskip \abovedisplayskip
\abovedisplayshortskip \z@ plus3pt\belowdisplayshortskip 6pt plus3pt
minus3pt\let\@listi\@listI}

\def\subsize{\@setsize\subsize{12pt}\xipt\@xipt}

\def\section{\@startsection {section}{1}{\z@}{24pt plus 2pt minus 2pt}
{12pt plus 2pt minus 2pt}{\large\bf}}

\def\subsection{\@startsection {subsection}{2}{\z@}{24pt plus 2pt minus 2pt}
{12pt plus 2pt minus 2pt}{\large\bf}}

\def\subsubsection{\@startsection {subsubsection}{3}{\z@}{24pt plus 2pt minus 2pt}
{12pt plus 2pt minus 2pt}{\large\bf}}

\makeatother

\title{\fontsize{16}{16}\selectfont Hierarchical modularity in human brain functional networks}

\begin{document}

\date{}

\maketitle

\noindent David Meunier$^{1,2}$, Renaud Lambiotte$^3$, Alex Fornito$^{1,2,4}$, Karen D. Ersche$^{1,2}$, Edward T. Bullmore$^{1,2,5}$\\ \\

\noindent1 Brain Mapping Unit, Department of Psychiatry, University of Cambridge, Cambridge, UK\\
2 Behavioural and Clinical Neurosciences Institute, University of Cambridge, Cambridge, UK\\
3 Institute for Mathematical Sciences, Imperial College, London, UK\\
4 Melbourne Neuropsychiatry Centre, Department of Psychiatry, University of Melbourne, Victoria, Australia \\
5 GSK Clinical Unit Cambridge, Addenbrooke's Hospital, Cambridge,
UK\\ \\

\noindent \textbf{Correspondence:}\\

\noindent Professor Ed Bullmore, Brain Mapping Unit, Herchel Smith
Building, Robinson Way, Cambridge, CB2 0SZ, UK\\

\noindent Email, etb23@cam.ac.uk. Fax, +44 (0)1223 336581. \\ \\

\noindent \textbf{Running title:}\\

\noindent Hierarchical modular human brain networks

\newpage

\section*{Abstract}
\noindent The idea that complex systems have a hierarchical modular
organization originated in the early 1960s and has recently
attracted fresh support from quantitative studies of large scale,
real-life networks. Here we investigate the hierarchical modular (or
``modules-within-modules'') decomposition of human brain functional
networks, measured using functional magnetic resonance imaging
(fMRI) in 18 healthy volunteers under no-task or resting conditions.
We used a customized template to extract networks with more than
1800 regional nodes, and we applied a fast algorithm to identify
nested modular structure at several hierarchical levels. We
 used mutual information, $0<I<1$, to estimate the similarity of community structure of
 networks in different subjects, and to identify the individual network that is most representative
of the group.
 Results show that human brain functional
 networks have a hierarchical modular organization with a fair degree of similarity between subjects, $I=0.63$. The largest 5 modules at the highest level of the hierarchy
 were medial occipital, lateral occipital, central, parieto-frontal and fronto-temporal systems; occipital modules
 demonstrated less sub-modular organization than modules comprising regions of multimodal association cortex. Connector nodes and hubs,
 with a key role in inter-modular connectivity, were also concentrated in association cortical areas.
 We conclude that methods are available for hierarchical modular decomposition of large numbers of
 high resolution brain functional networks using computationally expedient algorithms. This could enable future
 investigations of Simon's original hypothesis that hierarchy or near-decomposability of physical symbol
 systems is a critical design feature for their fast adaptivity to changing environmental conditions.     \\

\section*{Keywords}

\noindent graph theory, brain, network, modularity, hierarchy,
near-decomposability, information

\newpage

\section{Introduction}

Almost 50 years ago, Herbert Simon wrote an essay entitled ``The
architecture of complexity'' \citep{SIM62}. In this prescient
analysis, he argued that most complex systems, such as social,
biological and physical symbolic systems, are organized in a
hierarchical manner. He introduced the notion of
``nearly-decomposable systems'', i.e. systems where elements have
most of their interactions (of any kind) with a subset of elements
in some sense close to them, and much less interaction with elements
outside this subset. In mainstream contemporary parlance, Simon's
near-decomposability is closely analogous to the concept of
topological modularity: nodes in the same module have dense
intra-modular connectivity with each other and sparse inter-modular
connectivity with nodes in other modules \citep{NEW04b,NEW06}. Simon
argued that near-decomposability was a virtually universal property
of complex systems because it conferred a very important
evolutionary or adaptive advantage. Decomposability, or modularity,
accelerates the emergence of complex systems from simple systems by
providing stable intermediate forms (component modules) that allow
the system to adapt one module at a time without risking loss of
function in other, already-adapted modules.

 Our
understanding of complexity has progressed considerably since that
time, partly due to the availability of large data-sets that now
allow us to explore empirically the architecture of complex systems
and thereby to feedback on theoretical considerations
\citep{STR01,AMA04}. Many complex systems can be represented using
tools drawn from graph theory as networks of nodes linked by edges.
Such networks have been used to represent a broad variety of
systems, ranging from genetic and protein networks to the world wide
web. The huge size of some of these systems ($\sim$ 10 billion nodes
in the WWW) has driven the development of new statistical tools in
order to characterize their topological properties \citep{NEW03}.

A quantity called modularity has been introduced in order to measure
the decomposability of a network into modules
\citep{GUI04,NEW04}. Modularity can be used as a merit
function to find the optimal partition of a network. The resulting
partition has been shown to reveal important network community
structures in a variety of contexts, e.g., the global air
transportation network \citep{GUI05} and gene expression
interactomes \citep{OLD08} are two diverse examples of complex
systems with topological modularity. However, in systems having an
intrinsic hierarchical structure, finding a single partition is not
satisfactory. Several approaches have therefore been proposed in
order to allow for more flexibility and to uncover communities at
different hierarchical levels. Among those multi-scale approaches,
there are algorithms searching for local minima of the modularity
landscape \citep{SAL07} or modifying the adjacency matrix of the
graph in order to change its typical scale \citep{ARE08}. Another
class of methods consists in modifying modularity by incorporating
in it a resolution parameter \citep{REI06}. This allows one to
``zoom in and out'' of a modular hierarchy in order to find
communities on different levels; for example, the resolution
parameter can be interpreted as the time scale
 of a dynamical process unfolding on a network \citep{LAM09}.

There is already strong evidence that brain networks have a modular
organization; see \citep{BUL09} for review. Some support comes from
non-human data, like the anatomical networks in felines and primates
\citep{HIL00} or functional networks in rodents \citep{SCH08}.
Recently, human neuroimaging studies have also provided evidence for
comparable modular organization in both anatomical \citep{CHE08} and
functional \citep{FER08,MEU09} brain networks. However, a limitation
of these previous neuroimaging studies has been the computational
time required to derive a modular decomposition \citep{BRA06}, thus
limiting the size of the networks under study. In addition, these
studies were limited to studying modularity at one particular level
of community structure, neglecting consideration of possible
sub-modular communities at lower levels. Finally, it has been a
taxing problem to  quantify the topological similarity between two
or more modular decompositions, with most investigators simply
examining modularity on the basis of an averaged connectivity matrix
estimated from a group of individuals.

In this study,  we report on progress towards addressing each of
these issues. We applied a recently developed, computationally
efficient algorithm \citep{BLO08} to derive a hierarchical, modular
decomposition of human brain networks measured using fMRI in 18
healthy volunteers. By providing rapid decomposition, the algorithm
enabled us to study the modular structure of whole brain networks on
a larger scale (thousands of equally sized nodes) than previously
possible (tens of differently sized nodes), with concomitant
improvements in the spatial or anatomical resolution of the network,
while simultaneously avoiding biases associated with using {\em a
priori} anatomical templates that are inevitably somewhat arbitrary
in their definition of regions-of-interest \citep{TZO02}. Thus, the
method enabled rapid, high-resolution, hierarchical modular
decomposition of brain functional networks constructed from
individual fMRI datasets. In addition, we present a method for
comparing the similarity or mutual information between two modular
community structures obtained for different subjects, and use it to
identify the single, ``most representative'' subject whose brain
network modularity was most similar to that of all the other
networks in a sample of 18 healthy participants.

\newpage
\section{Material and Methods}

\subsection{Experimental data}

\subsubsection{Study sample}

Eighteen right-handed healthy volunteers (15 male, 3 female) were
recruited from the GlaxoSmithKline (GSK) Clinical Unit Cambridge, a
clinical research facility in Addenbrooke's Hospital, Cambridge, UK.
All volunteers (mean age: 32.7 years $\pm$ 6.9 SD) had a
satisfactory medical examination prior to study enrolment and were
screened for any other current Axis I psychiatric disorder using the
Structured Clinical Interview for the DSM-IV-TR Axis I Disorders
(SCID). Participants were also screened for normal radiological
appearance of structural MRI scans by a consultant neuroradiologist,
and female participants were screened for pregnancy. Urine samples
were used to confirm abstinence from illicit drugs and breath was
analyzed to ensure that no participant was under the influence of
acute alcohol intoxication. All volunteers provided written informed
consent and received monetary compensation for participation. The
study was reviewed and approved by the Cambridge Local Research
Ethics Committee (REC06/Q0108/130; PI: TW Robbins).

\subsubsection{Functional MRI data acquisition}

Whole-brain echoplanar imaging (EPI) data depicting BOLD contrast
were acquired at the Wolfson Brain Imaging Centre, University of
Cambridge, UK, using a Siemens Magnetom Tim Trio whole body scanner
operating at 3 Tesla with a birdcage head transmit/receive coil.
Gradient-echo, echoplanar imaging (EPI) data were acquired for the
whole brain with the following parameters: repetition time (TR) =
2000 ms; echo time (TE) = 30 ms, flip angle (FA) = 78 degrees, slice
thickness = 3 mm plus 0.75 mm interslice gap, 32 slices parallel to
the inter-commissural (AC-PC) line, image matrix size = 64 $\times$
64, within-plane voxel dimensions = 3.0 mm $\times$ 3.0 mm.

Participants were asked to lie quietly in the scanner with
eyes closed during the acquisition of 300 images. The first 4 EPI
images were discarded to account for T1 equilibration effects,
resulting in a series of 296 images, of which the first 256 images
were used to estimate wavelet correlations.

\subsubsection{Functional MRI data preprocessing}

\begin{figure*}[!h]
\includegraphics[width=6.0in]{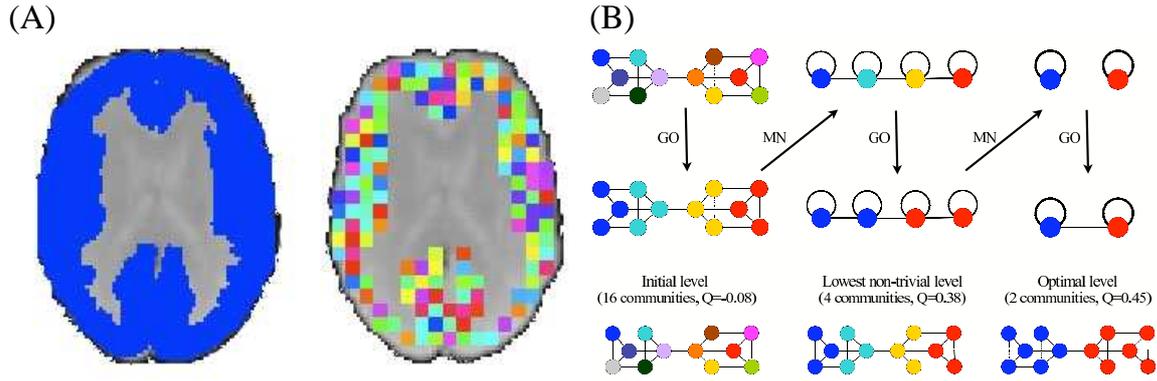}
\caption{Methods. (A) Downsampled template. Starting from a binary version of the AAL template (left), the downsampling procedure will produce a template of small (64 voxels), equal size regions covering the original template (right). (B) Illustration of the Louvain method on a simple hierarchical graph. The algorithm starts by assigning a different module to each node (16 modules of single nodes). The method then consists of two phases that are repeated iteratively. The first phase is a greedy optimisation (GO) where nodes adopt the community of one of their neighbours if this action results in an increase of modularity (typically, the community of the neighbour for which the increase is maximal is chosen). The second phase builds a meta-network (MN) whose nodes are the communities found in the first phase. We denote by ``pass" a combination of these two phases. Thepasses are repeated until no improvement of modularity is possible and the optimal partition is found. When applied on this graph, the algorithm first finds a lowest non-trivial level made of 4 communities. The next level is the optimal level and is made of 2 communities.}
\label{fig1}
\end{figure*}

The images were corrected for motion and registered to the standard
stereotactic space of the Montreal Neurological Institute EPI
template image using an affine transform \citep{SUC06}. Time series
were then extracted using a whole brain, high resolution, regional
parcellation of the images, implemented in the following manner; see
Figure \ref{fig1}(A). First, a binarized version of a commonly used
template image \citep{TZO02} was used as a broad gray matter mask.
Second, each $8
 \mathrm{mm}^3$ voxel in this mask was downsampled by a factor of 4
such that each equally sized region in the parcellation comprised
$4 \times 4 \times 4$ voxels of the original image.
This initial parcellation included some regions of the image which
were not largely representative of gray matter: these were excluded
from further analysis by applying the criteria that each region must
be at least 50\% overlapping with the grey matter mask and must
contain at least 80\% voxels having BOLD signal (defined
operationally as mean signal intensity greater than 50). To be
included in the definitive parcellation scheme (which comprised 1808
regional nodes), a region had to satisfy these two inclusion
criteria for every individual dataset in the sample.

The mean time series of each region was extracted and
wavelet-filtered using Brainwaver R package; \citep{ACH06,ACH07}
(\textit{http://cran.r-project.org/web/packages/brainwaver/index.html}).
The wavelet correlation coefficient was estimated for each of 4
wavelet scales between each pair of nodes, resulting in a $\{1808
\times 1808\}$ association matrix, or frequency-dependent functional
connectivity matrix, for each wavelet scale in the overall frequency
range 0.25-0.015 Hz. In what follows, we will focus on results at wavelet scale 3, subtending a frequency interval of 0.06-0.03 Hz.

This choice of frequency interval was guided by the fact
that prior work on resting-state fMRI functional connectivity has
found that the greatest power in connectivity occurs in frequency
bands lower than 0.1 Hz \citep{COR01}. However, analysing
very low frequency scales in limited time series such as those
acquired with fMRI can reduce precision in estimating inter-regional
wavelet correlations \citep{ACH06}. So scale 3 was chosen for the
focus of this study as representing a reasonable compromise between
retaining sufficient estimation precision while measuring low
frequency network properties.

Each association matrix was thresholded to create an adjacency
matrix $A$, the $a_{i,j}$th element of which is either 1, if the
absolute value of the wavelet correlation between nodes $i$ and $j$,
$w_{i,j}$, exceeds a threshold value $\tau$; or 0, if it does not.
We have chosen here to take a high threshold, leading to very sparse
networks comprising 8,000 edges, i.e., with a connection density of
0.5\% of all possible edges in a network of this size. Modularity of
neuroimaging networks is typically greater \citep{MEU09}, and
computational costs are lower, when the networks are more sparsely
thresholded. Up to 10\% of nodes were disconnected from the rest of
the network at this threshold.

\subsection{Graph theoretical analysis}

\subsubsection{Hierarchical modularity}

In recent years, many methods have been proposed to discover the
modular organization of complex networks. A key step was taken when
Girvan and Newman popularized graph-partitioning problems by
introducing the concept of modularity. Modularity is by far the most
widespread quantity for measuring the quality of a partition $\mathcal{P}$ of a
network. In its original definition, an unweighted and undirected
network that has been partitioned into communities has modularity
\citep{NEW04}:

\begin{eqnarray}
Q = \frac{1}{2m}\sum_{C \in \mathcal{P}} \sum_{i,j \in C}
\left[ A_{ij} - \frac{k_i k_j}{2m} \right] \,
\label{modAdef}
\end{eqnarray}

where  $A$ is the adjacency matrix of the network; $m$ is the total
number of edges; and $k_i = \sum_{j} A_{ij}$ is the degree of node
$i$. The indices $i$ and $j$ run over the $N$ nodes of the graph.
The index $C$ runs over the modules of the partition $\mathcal{P}$.
Modularity counts the number of edges between all pairs of nodes
belonging to the same community or module, and compares it to the
expected number of such edges for an equivalent random graph.
Modularity therefore evaluates how well a given partition
concentrates the edges within the modules.

A popular method for discovering the modules of a network consists
in optimizing modularity, namely in finding the partition having the
largest value of $Q$. However, it is typically impossible
computationally to sample modularity exhaustively by enumerating all
the possible partitions of a network into communities. Several
heuristic algorithms have therefore been proposed to provide good
approximations, and so to allow for the analysis of large networks
in reasonable times. The computational expediency of the algorithm
has become a crucial factor due to the increasing size of the
networks to be analyzed.

More recently, methods to study hierarchical modularity, also called
nested modularity, have been introduced \citep{SAL07,ARE08,ROS08}.
In this case, each module obtained at the partition of the highest
level can further be decomposed into submodules, which in turn can
be decomposed into sub-submodules, and so on. Here, we will use a
multi-level method which was introduced very recently in order to
optimize modularity \citep{BLO08}; see Figure \ref{fig1}(B). The
primary advantages of this method are that it unfolds a complete
hierarchical community structure for the network and outperforms
previous methods with respect to computation time. This so-called ``Louvain method'' takes advantage of the hierarchical organization of complex networks in order to facilitate the optimization. The algorithm starts by assigning a different module to each node of the network. The initial partition of the network is therefore made of $N$ communities. It then consists of two phases that are repeated iteratively. The first phase consists in a greedy optimization where nodes are selected sequentially in an order that has been randomly assigned. When a node is selected, it may leave its community and adopt a community which is in its direct neighbourhood, but only if this change of community leads to an increase of modularity (GO on Figure \ref{fig1}(B)). The second phase builds a new network whose meta-nodes are the communities found in the first phase (MN on Figure \ref{fig1}(B)). Let us denote by ``pass'' a combination of these two phases. These passes
are repeated iteratively until a maximum of modularity is attained
and an optimal partition of the network into communities is found.
By construction, the meta-nodes, or intermediate communities, are
made of more nodes at subsequent passes. The optimization is
therefore done in a multi-scale way: among adjacent nodes at the
first pass, among adjacent meta-nodes at the second pass, etc. The
output of the algorithm is a set of partitions, one for each pass.
The optimal partition is the one found at the last pass. It has been
shown on several examples that modularity estimated by this method
is very close to the optimal value obtained from slower methods
\citep{BLO08}. Intermediate partitions can also be shown to be
 meaningful and to correspond to communities at intermediate
resolutions (see Section \ref{issues}). In the following, we will call ``lowest non-trivial
level" the partition found after the first pass.

\subsubsection{Node roles}

Once a maximally modular partition of the network has been
identified, it is possible to assign topological roles to each node
based on its density of intra- and inter-modular connections
\citep{GUI05,GUI05b,GUI05c,SAL07}.

Intra-modular connectivity is measured by the normalized
within-module degree,
\begin{equation}
z_i \ = \ \frac{\kappa_{n_i}
-\overline{\kappa}_{n}}{\sigma_{\kappa_n}}
\end{equation}

\noindent where $\kappa_{n_i}$ is the number of edges connecting the
$i^{th}$ node to other nodes in the $n^{ th}$ module,
$\overline{\kappa}_n$ is the average of $\kappa_{n_i}$ over all
nodes in the module $n$, and $\sigma_{\kappa_n}$ is the standard
deviation of the intra-modular degrees in the $n^{th}$ module. Thus
$z_i$ will be large for  a node that has a large number of
intra-modular connections.

Inter-modular connectivity is measured by the participation coefficient,
\begin{equation}
P_i \ = \ 1 - \sum_{n=1}^{N} \left( \frac{\kappa_{n_i}}{k_i}
\right)^2
\end{equation}

\noindent where $\kappa_{n_i}$ is the number of edges linking the
$i^{th}$ node to other nodes in the $n^{th}$ module, and $k_i$ is
the total degree of the $i^{th}$ node. Thus $P_i$ will be close to
one if the $i^{th}$ node is extensively linked to all other modules
in the community and zero if it is linked exclusively to other nodes
in its own module.

The two-dimensional space defined by these parameters, the $\{P,z\}$
plane, can be partitioned to assign categorical roles to the nodes
of the network. Contrarily to our previous study \citep{MEU09},
where we used a simplified definition of node roles, the higher
number of nodes examined in the current study allowed us to adopt
the original definitions of node roles as described for large
metabolic \citep{GUI05} and transportation networks \citep{GUI05b}:

\begin{itemize}
\item The hubness of a node can be defined by its within-module degree:
If a given node $i$ has a value of $z_i > 2.5$. it is classified as
a hub, otherwise as a non-hub.
\item The limits for the participation coefficient are different for hubs and non-hubs.
For non-hubs, if a given node has value $0 < P_i < 0.05$, the node
is classified as an \textit{ultra-peripheral node}, $0.05 < P_i <
0.62$ corresponds to a \textit{peripheral node}, $0.62 < P_i < 0.80$
corresponds to a \textit{connector node}, and $0.80 < P_i < 1.0$ is
a \textit{kinless node}. For hubs, $0 < P_i < 0.30$ corresponds to a
a \textit{provincial hub}, $0.30 < P_i < 0.75$ corresponds to a
\textit{connector hub}, and $0.75 < P_i < 1.0$ is a \textit{kinless
hub}.
\end{itemize}

These different categories allowed us to classify the nodes
according to their topological functions in the network.
 For example, a provincial hub is a hub with greater intra- vs inter-modular connectivity, thus having a
 pivotal role in the function realized by its module, whereas a connector hub will play a central role
 in transferring information from its module to the rest of the network.

The results of modular decomposition were visualised in anatomical
space using Caret software for cortical surface mapping
(http://brainmap.wustl.edu/register.html), and in topological space
using Pajek software
(http://vlado.fmf.uni-lj.si/pub/networks/pajek).

\subsubsection{Similarity measure}

To compare the different modularity partitions obtained at different
hierarchical levels in the same subject, or at the same hierarchical
level in different subjects, we used the normalized mutual
information, as defined in \cite{KUN04}. For two given partitions
$A$ and $B$, with a number of communities denoted $C_A$ and $C_B$:

\begin{equation}
I(A,B)=\frac{-2\sum^{C_A}_{i=1}\sum^{C_B}_{j=1}
N_{ij}\log\left(\frac{N_{ij}N}{N_{i.}N_{.j}}\right)}
{\sum^{C_A}_{i=1}N_{i.}\log\left(\frac{N_{i.}}{N}\right)
 + \sum^{C_B}_{j=1}N_{.j}\log\left(\frac{N_{.j}}{N}\right)}
\label{eq:sim}
\end{equation}

where $N_{ij}$ is the number of nodes in common between modules $i$
and $j$, the sum over row $i$ of matrix $N_{ij}$ is denoted
$N_{i.}$, and the sum over column $j$ is denoted $N_{.j}$. If the
two partitions are identical then $I(A,B)$ takes its maximum value
of 1. If the two partitions are totally independent, $I(A,B)= 0$.

The initial application of this quantity was to evaluate different
modularity partition algorithms \citep{DAN05}. The similarity index
was used to compute how closely the partitions obtained from
different algorithms matched the ``target'' partition of a given
test network, i.e., a network whose modular structure was known
\textit{a priori}. Here the application was different, since we
wanted to compare partitions obtained for different subjects in a
group. Since the equation is symmetric in $A$ and $B$, it is however
possible to use the index without a target partition.

The networks constructed for each individual had the same number of
nodes $N$, so the partitions of each subject have the same number of
nodes. However, due to the high threshold applied to construct the
adjacency matrix, the number of disconnected nodes in the networks
can be different for each subject. One solution is to consider each
disconnected node as a single module. In this case, each node
(disconnected or not) of the network will be in the set of modules
of each subject. However, it introduces artificially high values in
the similarity values, especially if the networks of two subjects
have similar sets of disconnected edges. So we have chosen to remove
the disconnected nodes from the partitions and study only the
partitions obtained on the giant component of each network, but
keeping the value of $N$ in the equation as the total number of
nodes. This leads to a value of similarity slightly lower than if
the disconnected nodes were included in the partitions, but is more
representative of the relevant set of connected modules.

\newpage
\section{Results}

\subsection{Similarity and variability of modular decompositions}
\label{simil}

\begin{figure*}[!h]
\includegraphics[width=6.0in]{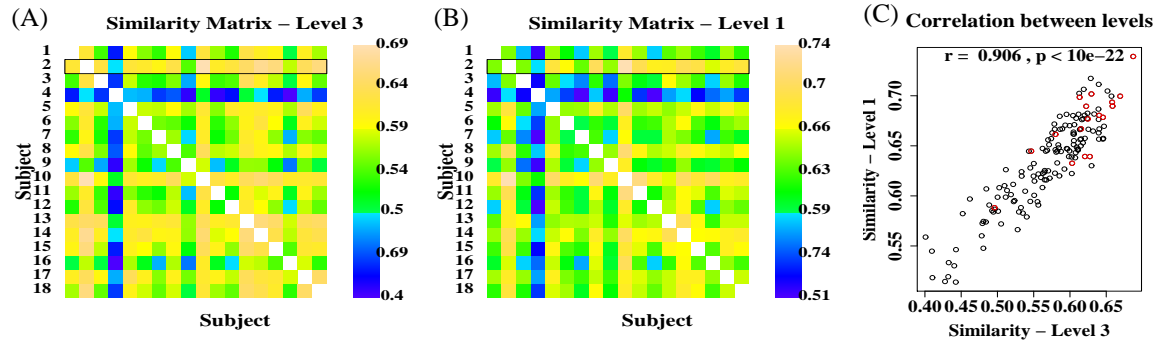}
\caption{Variability and similarity of brain functional network
community structure between 18 different subjects. (A) Matrix showing
the between-subject similarity measure for community structure at
the highest level of the modular hierarchy. The pairwise similarity
scores for the most representative subject are highlighted by a
black rectangle. (B) Matrix showing the between-subject similarities
for community structure at the lowest level of the modular
hierarchy. (C) Scatterplot showing strong correlation of
between-subject similarities at high and low levels of the modular
hierarchy. Red points are similarities for the most representative
subject.} \label{fig2}
\end{figure*}

It was possible to define a hierarchical modular decomposition for
each of the 18 subjects in the sample. At the highest hierarchical
level, the mean brain functional network modularity for the group
was 0.604, with standard deviation (SD) = 0.097. By comparison,
modularity at the highest level for 18 random networks with an
equivalent number of nodes (1808) and edges (8000) was 0.303 (SD =
0.003). There was a significant increase in brain network modularity
compared to random network modularity (Kolmogorov-Smirnov test, $D =
1 , P \sim 2^{-10}$).

The similarity of network
community structure between each pair of subjects, at each level of
the hierarchy, was calculated using Eq \ref{eq:sim}. The resulting
similarity matrices for level 3 (the highest level) and level 1 (the
lowest non-trivial level) are shown in Figure \ref{fig2}.

The average pairwise similarity was 0.57 at level 3 and 0.63 at level 1,
indicating a reasonable degree of consistency between subjects in
modular organization of functional networks. The similarity between
subjects was highly correlated over levels of the modular hierarchy:
for example, if a pair of networks had a similar modular partition
at the highest level, the submodular organization at lower levels
was also similar.

Simply by summing the pairwise similarity scores for each row of the
similarity matrix, it was possible to identify the individual
subject (number 2) that was most similar to all other subjects in
the sample, i.e., the most representative subject, and the subject
(4) that was least similar to the rest of the sample. In what
follows, we will focus attention on the modular decomposition of the
most representative subject.

\subsection{Hierarchical modularity}

\begin{figure*}[!h]
\includegraphics[width=6.0in]{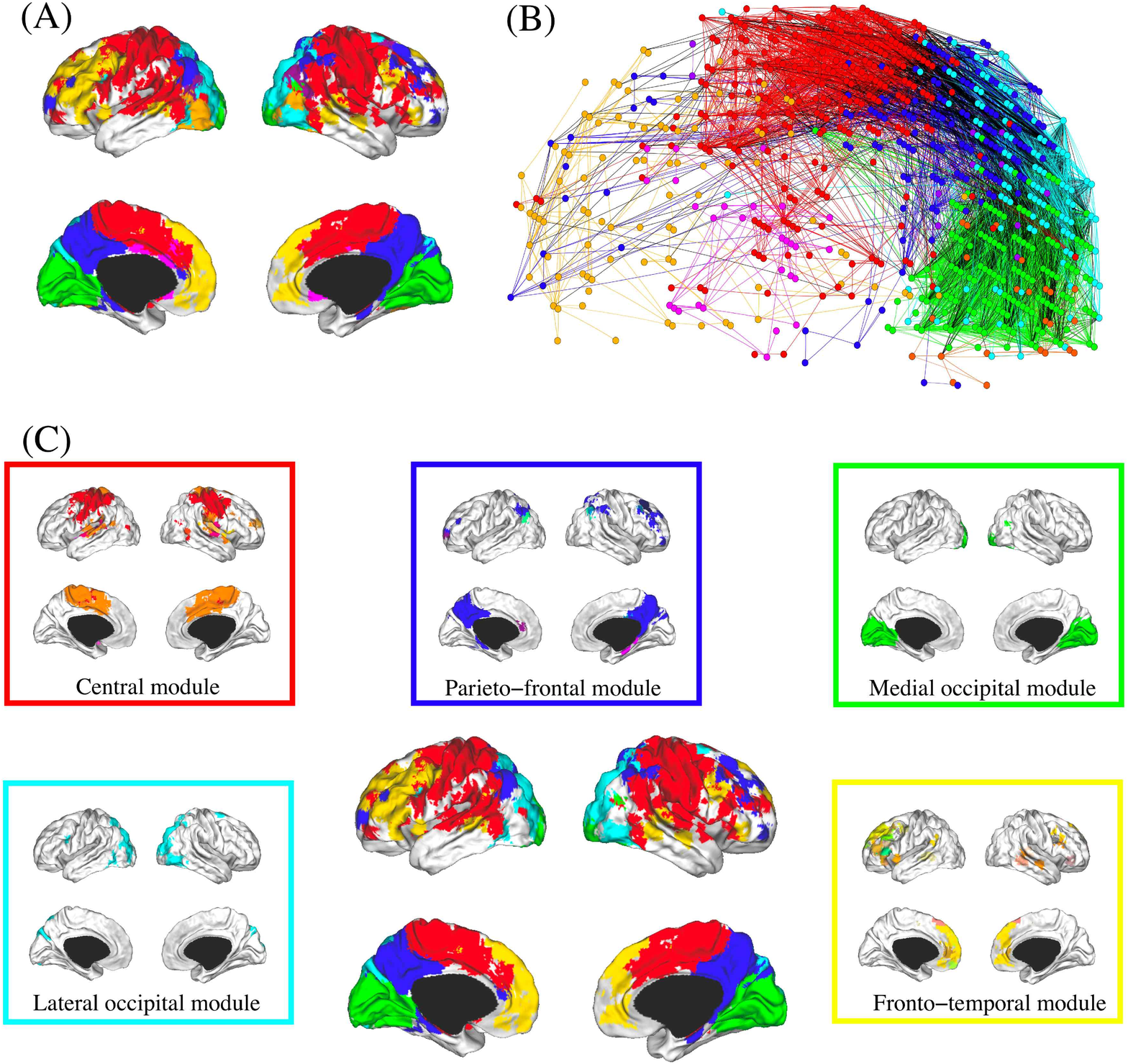}
\caption{Hierarchical modularity of a human brain functional
network. (A) Cortical surface mapping of the community structure of
the network at the highest hierarchical level of modularity, showing
all modules that comprise more than 10 nodes.  (B) Anatomical
representation of the connectivity between nodes in color-coded
modules. The brain is viewed from the left side with the frontal
cortex on the left of the panel and the occipital cortex on the
right of the panel. Intra-modular edges are colored differently for
each module; inter-modular edges are drawn in black. (C) Sub-modular
decomposition of the five largest modules (shown centrally)
illustrates that the medial occipital module has no major
sub-modules whereas the fronto-temporal modules has many
sub-modules.} \label{fig3}
\end{figure*}

The hierarchical modular decomposition of the most representative
subject's brain functional network is shown in Figure \ref{fig3}. At
the highest level of the hierarchy (level 3), there were 8 large
modules, each comprising more than 10 nodes. At the lowest level of
the hierarchy (level 1), there were 57 sub-modules. The largest 5
modules (with putative functional interpretations) and their
sub-modular decomposition are briefly described below; some
additional details are provided in Table \ref{tab1}.

\begin{table}
\resizebox{6.5in}{!}
{
\begin{tabular}{|c|c|c|c|c|c|c|c|} \hline
Module description & \# Nodes &  Connector Nodes & Provincial
Hubs & Connector Hubs & Submodules & Size of submodules \\  \hline
\hline Central (sensorimotor) & 239 &
8 & 1 & 4 & 11 & 115, 96, 8, 4, 3 \textit{(2)}, 2 \textit{(5)}\\
\hline Parieto-frontal(default/attention) & 138 & 10 & 1 & 0 & 10 &
115, 3 \textit{(5)}, 2 \textit{(4)}\\ \hline Medial occipital
(primary visual) & 132 & 3 & 0 & 0 & 1 & 132\\ \hline Lateral
occipital (secondary visual) & 101 & 7 & 0 & 1 & 1 & 101\\ \hline
Fronto-temporal (symbolic) & 89 & 0 & 2 & 3 & 24 & 19, 8, 6, 5
\textit{(2)}, 4, 3 \textit{(6)}, 2 \textit{(12)}\\ \hline
\end{tabular}
}

\caption{\label{tab1} The five largest modules of the human brain
functional network in a representative normal volunteer, indicating
the number and type of nodes and submodules.}
\end{table}

\begin{itemize}
\item Central module (somatosensorimotor): The largest high level module comprised
extensive areas of lateral cortex in premotor, precentral and
postcentral areas, extending inferiorly to superior temporal gyrus,
as well as to premotor and dorsal cingulate cortex medially. At a
lower hierarchical level, medial and lateral cortex were segregated
in different sub-modules and, within lateral cortex, precentral and
postcentral areas were segregated from superior temporal cortex.
\item Parieto-frontal module (default/attentional): This module comprised
medial posterior parietal and posterior cingulate cortex, extending
to medial temporal lobe structures inferiorly, and areas of inferior
parietal and dorsal prefrontal cortex laterally.
\item Medial occipital module (primary visual): This module
comprised medial occipital cortex and occipital pole, including
primary visual areas.
\item Lateral occipital (secondary visual): This module comprised
dorsal and ventral areas of lateral occipital cortex, including
secondary visual areas.
\item Fronto-temporal module (symbolic): This module comprised dorsal and
ventral lateral prefrontal cortex, medial prefrontal cortex, and
areas of superior temporal cortex. It was less symmetrically
organized than most of the other high level modules and was
decomposed to a larger number of sub-modules at lower levels.
\end{itemize}

Note that most high level modules are bilaterally symmetrical,
comprise both lateral and medial cortical areas, and tend to be
spatially concentrated in an anatomical neighborhood. Submodular
decomposition sometimes resulted in a dominant sub-module,
comprising most of the nodes in the higher level module, with some
much smaller sub-modules each comprising a few peripheral nodes. For
example, this was the pattern for the occipital modules. An
alternative result was a more even-handed decomposition of a high
level module into multiple sub-modules; this was the pattern for the
prefronto-temporal module. In Simon's terminology, the number of
sub-modules into which a module can be decomposed is its span of
control, and so we can describe occipital modules as having a
greater span of control than, say, the fronto-temporal module.

\subsection{Node roles}

\begin{figure*}[!h]
\includegraphics[width=6.0in]{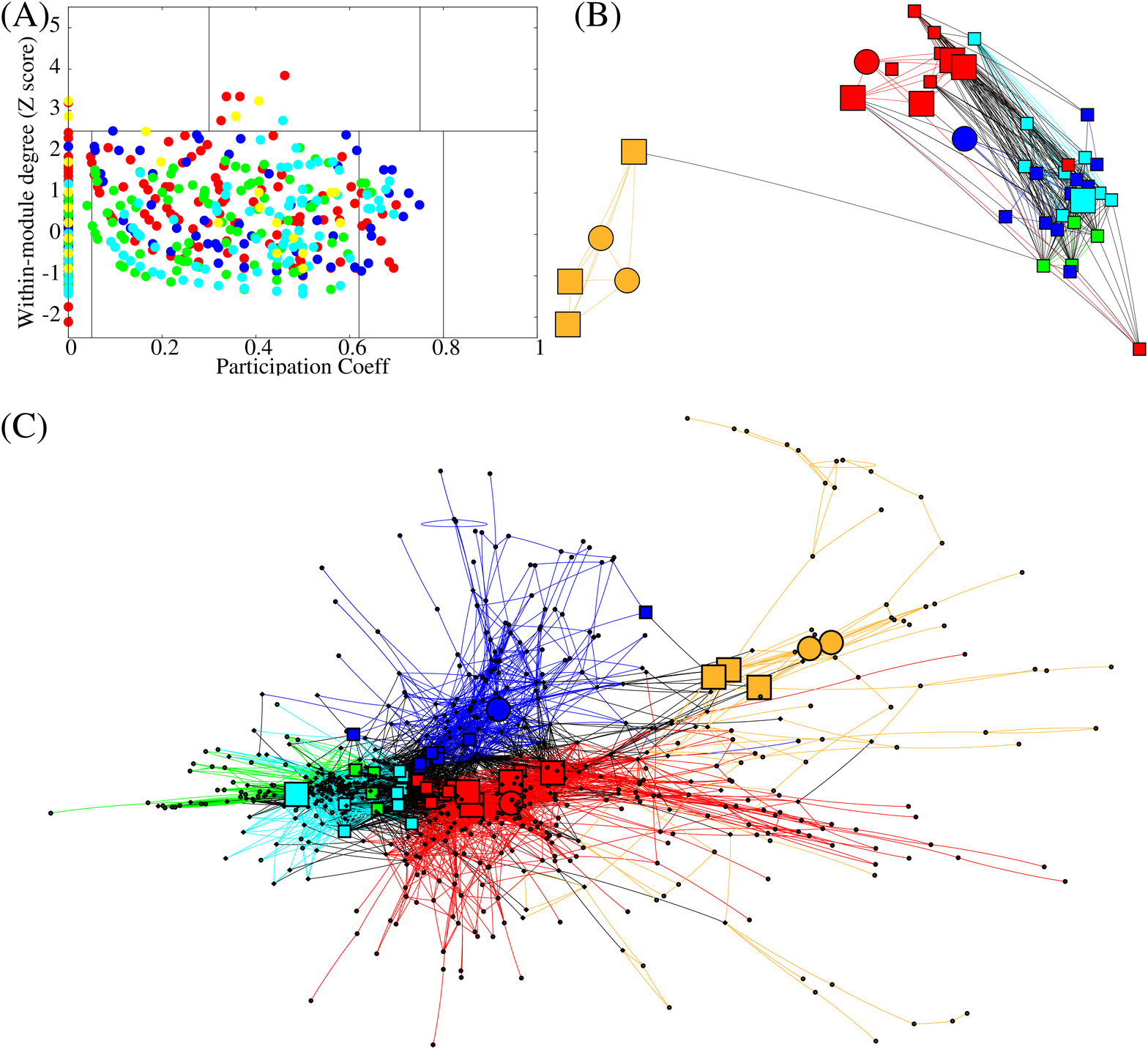}
\caption{Toplogical roles of network nodes in intra- and
inter-modular connectivity. (A) All nodes are plotted in the
$\{P-z\}$ plane of intra-modular degree $z$ versus participation
coefficient $P$; the solid lines partition the plane according to
criteria for hubs versus non-hubs and connector, provincial,
peripheral or kinless nodes. (B) Anatomical representation of the
provincial hubs (circles), connector hubs (large squares) and
connector nodes (small squares) of each of each of the 5 largest
modules at the highest level of the modular hierarchy. (C)
Topological representation of the network in using
Fructerman-Reingold algorithm \citep{FRU91} to highlight topological
proximity of highly connected nodes; color and shape of the nodes
represent their modular assignment and topological role as above and
in Figure 2. }
\label{fig4}
\end{figure*}

On the basis of the highest level (level 3) of modular decomposition, we assigned
topological roles to each of the regional nodes. A node was defined
as a hub or non-hub (more or less highly connected) with a
provincial, connector or kinless role (depending on its balance of
intra- vs inter-module connectivity). Provincial hubs will play a
key role in intra-modular processing; connector hubs will play a key
role in inter-modular processing.

Figure \ref{fig4} displays an example of the node roles obtained
from the most representative subject. Figure \ref{fig4}(A) shows the
participation coefficient ($P$, our measure of inter-modular
connectivity) {\em vs} the intra-modular degree ($z$, our measure of
hubness) for each regional node in the network. Most nodes (416, 53\%) have no
inter-modular connections $P=0$, but some (28,4\%) have a high proportion of inter-modular
connections, qualifying for connector status. Figure \ref{fig4}(B)
is a spatial representation of the node roles, the locations of the
nodes corresponding to their position in three-dimensional
stereotactic space. Figure \ref{fig4}(C) is a topological
representation obtained by applying the Fruchterman-Reingold
algorithm \citep{FRU91} to the network displayed in Figure
\ref{fig4}(B). In this representation, the distances between the
nodes are not related to their spatial location, but to how strongly
linked connected they are to their neighbours. The main idea is to
start from an initial random placement of the nodes, and replace the
edges by springs, letting the equivalent mechanical system evolve
until it reaches a stable mechanical state. Thus, this
representation locates nodes with similar connectivity patterns
closer together in space.

We can see that most nodes (743, i.e. 95\% of the nodes) have either the role of ultra-peripheral nodes or
peripheral nodes and a small minority (39, i.e. 5\% of the nodes) have the topologically important roles of
hubs and/or connector status. Inter-modular connections, and the
connector nodes and hubs which mediate them, are most numerous in
posterior modules containing regions of association cortex; the
fronto-temporal module is sparsely connected to other modules and
the medial occipital module also has relatively few connector nodes.

\subsection{Methodological issues}
\label{issues}

\begin{figure*}[!h]
\includegraphics[width=6.0in]{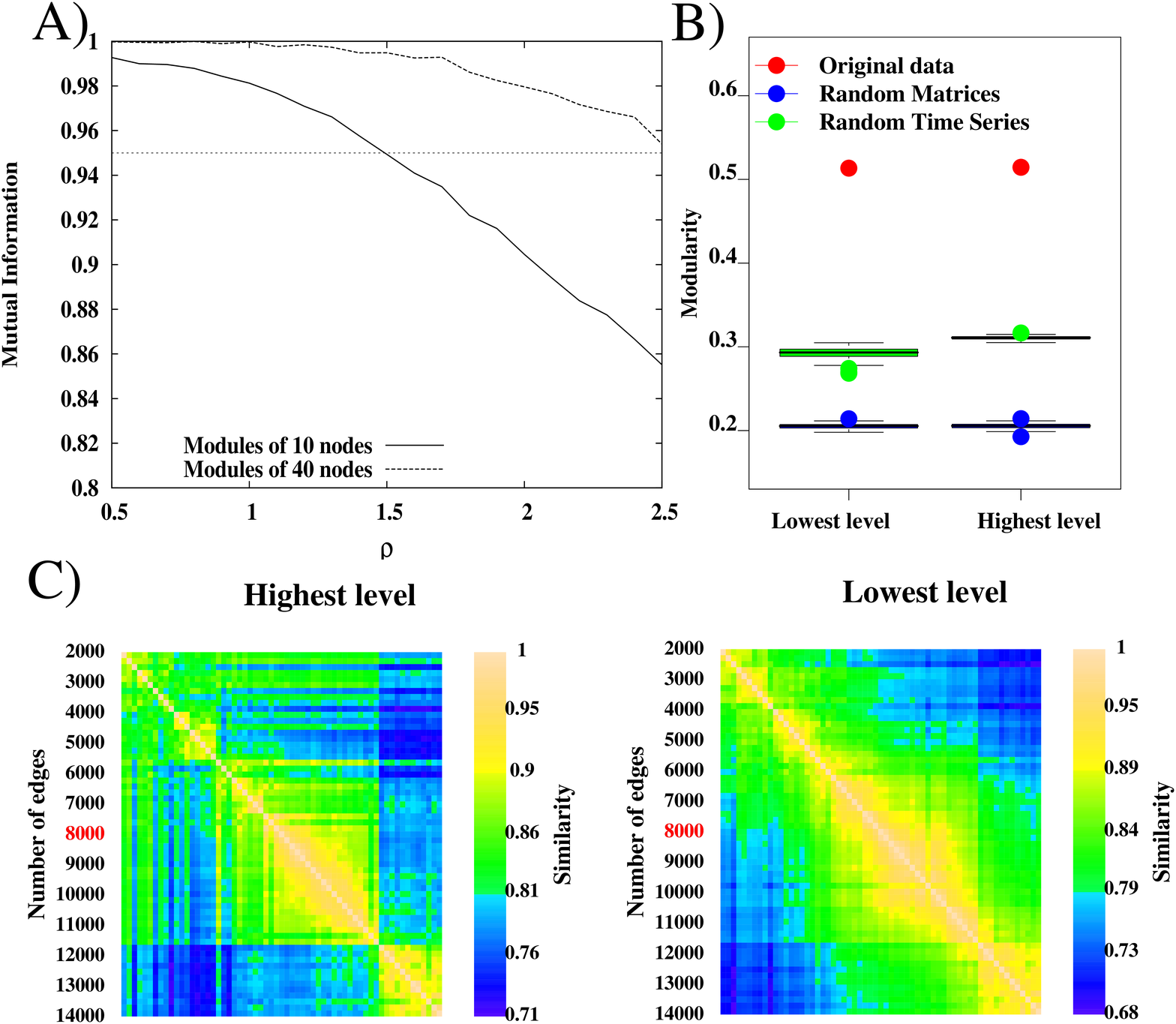}
\caption{Methodological issues in analysis of hierarchical
modularity. (A) Validation of the Louvain method for
hierarchical decomposition on a benchmark network defined in
(Sales-Pardo et al., 2007). The network is naturally made of 64, 16
and 4 modules of 10, 40 and 160 nodes respectively. The separability
of different levels of the benchmark network is controlled by the
parameter $\rho$. We calculate the normalized information between
the lowest non-trivial level partition and the natural partition of
64 modules (solid curve), and between the second level partition and
the natural partition of 16 modules (dashed curve). After averaging
over 20 different realizations of the network, our simulations show
an excellent agreement as mutual information is above 0.95 for
values of $\rho$ up to 1.5 for the lowest non-trivial and
intermediate levels. (B) Modularity values at the highest and
lowest levels of hierarchical community structure in a
representative brain network (Subject ID 2, in red) and for networks
obtained from 100 randomizations of the original time-series (in
green), and for networks obtained by 100 randomizations of the
original adjacency matrix. (C) Similarity measures between highest
level partitions (left) and non-trivial lowest level partitions
(right) obtained by thresholding the original network to retain
different number of highest correlations as edges.} \label{fig5}
\end{figure*}

This work is a first attempt to uncover the hierarchical organisation of brain functional networks and to compare the stability of hierarchical modular decompositions across individuals. There are, however, three possible weaknesses in our analysis that we would like to address in this section.

\subsubsection{Validation of the algorithm}

A first consideration concerns the choice of the Louvain method (LM) in order to uncover nested modules in the brain networks. LM was first proposed in order to uncover optimal partitions of a graph by maximising modularity. This is a greedy method which is known to be very fast and very precise \citep{BLO08}, albeit less precise than much slower methods such as Simulated Annealing (SA). It is interesting to note, however, that this lack of precision may be an advantage, in practice, as it may avoid some of the pitfalls of modularity analysis such as its resolution limit \citep{FOR07}. For instance, it has been recently shown that LM performs much better than SA when applied to benchmark networks with unbalanced modules comprising different numbers of nodes \citep{LAN09}. We therefore believe that there is good evidence that the top level partitions uncovered by LM are valid. The validity of the intermediate hierarchical levels identified by the algorithm is, however, more arguable, as it has not been studied in detail yet \citep{BLO09}. In order to show the validity of these intermediate levels, we need to verify that the method uncovers all the significant partitions present in the network and only those.

To do so, we have tested LM on a benchmark network with known hierarchical structure (\citealp{SAL07}; Figure \ref{fig5}(A)). This benchmark network is made of 640 nodes with three levels of organisation: small modules comprising 10 nodes, medium-size modules comprising 40 nodes and large modules comprising 160 nodes. The cohesiveness of the hierarchy between levels is tuned by a single parameter $\rho$, i.e. the larger the value of $\rho$, the more difficult it is to find the sub-modules. When applied on this benchmark network, the algorithm finds with an excellent precision the first two levels (16 modules and 64 modules), but does not uncover the partition into 4 modules. This result is to be expected because this partition into 4 modules is sub-optimal in terms of modularity and can therefore not be uncovered by an aggregative method. This shows that the method can at best uncover the partition optimising modularity and finer partitions. In order to uncover coarser partitions, one needs to decrease the resolution of the method, which can be done by following \cite{REI06}, or \cite{SAL07}, for instance.

On the same benchmark network, the algorithm typically finds two
levels (one corresponding to 64 modules and one corresponding to 16
modules) but it may occasionally find three levels (one level
corresponding to 64 modules and two levels similar to the partition
into 16 modules). When $\rho=1.0$, for instance, over 100
realisations of the graph,  the algorithm finds two levels on 86
realisations, and three levels on 14 realisations. This result is
encouraging as it suggests that the algorithm only produces
significant partitions. However, it is possible to find situations
where it is not the case, e.g., random graphs. It is therefore still
necessary to verify the significance of intermediate partitions, as
we will discuss below.

 \subsubsection{Comparison with a random graph}

A second consideration concerns the comparison of the partition
of the original network with randomised data, as the algorithm also
gives a hierarchical decomposition for randomly generated networks.
To show that the representative brain network under study (subject
ID 2) displays a non-random hierarchical modular structure, we have
randomised the original data and processed the hierarchical
structure of randomised networks, with two kind of randomisation.
First, by computing 100 randomisations of the time points in the
original time-series (in green on Figure \ref{fig5}(B)) and, second,
by randomising the original adjacency matrix 100 times (in blue on
Figure \ref{fig5}(B)). Note that the two kinds of randomisation lead
to networks with different sizes: in the randomised time-series
networks, almost all the nodes are connected, thus leading to
networks with 1808 nodes and 8000 edges. Whereas starting from the
original adjacency matrix leads to networks of 844 nodes and 8000
edges. The modularity obtained for the lowest and highest partitions
of the original network are displayed in Figure \ref{fig5}(B). The
modularity values are clearly reduced in the randomized networks,
relative to the original data, indicating that our results on real
brain networks are not trivially reproduced in random networks.

In order to show that the intermediate levels considered in this
paper are significant, we have followed the argument that
significant partitions should be robust, in the sense that they
should only be weakly altered by a modification of the optimisation
algorithm. As argued by \cite{RON09}, comparing the optimal
partitions found by the algorithm for different orders of the nodes
is a way to test their robustness and therefore their validity. We
have therefore optimized the modularity of the representative brain
network 100 times by choosing the nodes in a different order, and
focused on the first non-trivial partition found by the algorithm.
The mutual information between pairs of partitions obtained for each
different order is then computed. The average mutual information
among those pairs is very high (0.89) compared to what is obtained
for a comparable random network (0.44), thereby suggesting that
partitions obtained at the lowest non-trivial levels are relevant
for the network under study.

\subsubsection{Dependence on the number of edges}

A third consideration concerns the number $m$ of edges that we have chosen in order to map the correlation matrices onto unweighted graphs. This is a known problem when dealing with fMRI data and building brain networks. If $m$ is too small, i.e. keeping the top most significant links, the network will be so sparsely connected that it will be made of several disconnected clusters. If $m$ is too large, in contrast, the network will be very densely connected, but mainly made of unsignificant links. In these two extremes, the network structure is a bad representation of the correlation matrix. This is still an open problem that requires the right trade-off between these two competing factors. In order to show the robustness of our results, we propose to look at the resilience of the hierarchical modular organisation under the tuning of the value of $m$. Meaningful values of $m$ are identified by intervals over which the structure of the network is preserved.  We have applied this scheme to the optimal partitions of the most representative subject (Subject 2), over a wide range of threshold (2000 edges to 14000 edges, with a step of 200 edges). Our results show that partitions are very similar (in terms of mutual information) over the range [6000-11000] for both highest level (left on Figure \ref{fig5}(C)) and non-trivial lowest level (right on Figure \ref{fig5}(C)), indicating our results are robust to the specific choice of threshold.

\newpage
\section{Discussion}

In this study, we have applied recently developed tools for
characterizing the hierarchical, modular structure of complex
systems to functional brain networks generated from human fMRI data
recorded under no-task or resting state conditions. Where previous
comparable work was limited by the computational expense of
available modularity algorithms, meaning that only one or a few
relatively low resolution networks (comprising 10s of nodes) could
be analysed, here we were able to obtain modular decompositions on a
larger number of higher resolution networks (each comprising 1000s
of nodes). In addition, we used an information-based measure to
quantify the similarity of community structure between two different
networks and so to find a principled way of focusing attention on a
single network that is representative of the group.

\subsection{Hierarchical modularity}

There was clear evidence for hierarchical modularity in these data
and the community structure of the networks at all levels of the
hierarchy was reasonably similar across subjects ($I \sim 0.6$),
suggesting that brain functional modularity is likely to be a
replicable phenomenon. This position is further supported by the
qualitative similarity between the major modules identified at the
highest level of the hierarchy in this study and the major modules
or functional clusters identified in comparable prior studies on
independent samples \citep{SAL05,MEU09}. As previously, the major functional modules
comprised functionally and/or anatomically related regions of cortex
and this pattern was also evident to some extent at sub-modular
levels of analysis. For example, the central module comprising areas
of somatosensorimotor and premotor cortex was segregated at a
sub-modular level into a medial component, comprising supplementary
motor area and cingulate motor area, and a lateral component,
comprising precentral and postcentral areas of primary motor and
somatosensory cortex.

Another plausible aspect of the results was the clear evidence for a
symmetrical posterior-to-anterior progression of cortical modules.
This was seen most clearly on the medial surfaces of the cerebral
hemispheres in terms of their division into medial occipital,
parieto-frontal and central modules. A posterior-to-anterior
organization of cortical modules in adult brain functional networks
is arguably compatible with the abundant evidence from
neurodevelopmental studies which have shown rostro-caudal modularity
of the spinal cord, brain stem, hind brain and diencephalon defined
by segmented patterns of gene expression \citep{RED01}. This
speculative link between the topological modularity of adult brain
networks and the embryonic modularity of the developing nervous
system presents an interesting focus for future studies.

\subsection{Node roles in inter-modular connectivity}

One important potential benefit of a modular analysis of complex
networks is that it allows us to be more precise about the
topological role of any particular node in the network. For example,
rather than simply saying that a particular region has a high degree
we may be able to say that it has a disproportionately important
role in transfer of information between modules, rather than within
a module. In these data, the location of connector nodes and hubs
with a prominent role in inter-modular communication was
concentrated in posterior areas of association cortex. The
fronto-temporal module, on the other hand, was rather sparsely
connected to other modules. One possible explanation for these
anatomical differences in inter-modular communication may relate to
the stationarity of functional connectivity between brain regions.
Our measure of association between brain regions (the wavelet
correlation corresponding to a frequency interval of 0.03-0.06 Hz)
provides an estimate of functional connectivity ``on average'' over
the entire period of observation (8 mins
35 s). If there is significant variability over time in the strength
of functional connections between modules this may be manifest in
terms of reduced connectivity on average over a prolonged period.
Thus one possible explanation for the sparser inter-modular
connections of the fronto-temporal module is that the interactions
of this system with the rest of the brain network may be more
non-stationary or labile over time. This interpretation could be
tested by future studies using time-varying measures of functional
connectivity, such as phase synchronization \citep{KIT09}.

\subsection{Dealing with more than one subject}

One of the challenges in analysis of network community structure is
the richness of the results (every node will have a modular
assignment and a topological role) and the difficulties attendant on
properly managing inter-individual variability in such novel
metrics. In previous work, we estimated a functional connectivity
matrix for each subject, then thresholded the group mean
association, and explored the community structure of the group mean
network. This allows us to focus attention on a single network but
it neglects between-subject variability and, like any use of the
mean in small samples, it is potentially biased by one or more
outlying values for the functional connectivity. Here we have
explored an alternative approach, using an information-based measure
of similarity to quantify between subject differences in network
organization and to identify the most representative subject in the
sample. One can imagine that this measure could be combined with
resampling based approaches to statistical inference in order to
estimate, for example, the probability that the community structure
identified in a single patient is significantly dissimilar to a
reference group of brain networks in normal volunteers. However, it
fair to say that there are a number of technical challenges to be
addressed in using modularity measures for statistical comparisons
between different groups.

\subsection{Returning to Simon's hypothesis}

As this is the first study to attempt a hierarchical modular decomposition of human brain functional networks, there is little guidance in the existing literature as to what the correct structure of the network should resemble. Our results are encouraging in that they have been able to identify well defined neuroanatomical systems, but they remain empirical and require further validation in appropriate animal models. However, our analysis of simulated data (section \ref{issues}) indicates that our algorithm does indeed identify the correct structure of a hierarchical, modular network, which lends confidence to our results.

In Simon's theoretical analysis, near-decomposability was considered
to be a ubiquitous property of complex systems because it conferred
advantages of adaptive speed in response to evolutionary selection
pressures as well as shorter-term developmental or environmental
contingencies. In relation to the modularity of human brain systems,
this view prompts a number of questions. Perhaps the most
immediately addressable, at least by functional neuroimaging, is the
question of how the modularity of brain network organization relates
to cognitive performance and the capacity to shift attention rapidly
between different stimuli or tasks. According to Simon's theory,
this key aspect of the brain's cognitive function should depend
critically on modular or sub-modular components and the rapid
reconfiguration of inter-modular connections between them. Future
studies, applying graph theoretical techniques to modularity
analysis of fMRI data recorded during task performance (rather than
in no-task state) may be important in testing this prediction.

\section{Conclusion}

We have described graph theoretical tools for analysis of
hierarchical modularity in human brain functional networks derived
from fMRI. Our main claims are that these techniques are
computationally feasible and generate plausible and reasonably
consistent descriptions of the brain functional network community
structure in a group of normal volunteers. The potential importance
theoretically of this analysis has been highlighted by reference to
Simon's seminal theory of hierarchy and decomposability in the
design of information processing systems.

\newpage
\section*{Acknowledgements}

This research was supported by a Human Brain Project grant from the
National Institute of Mental Health and the National Institute of
Biomedical Imaging \& Bioengineering, National Institutes of Health,
Bethesda, MD, USA. RL was supported by UK EPSRC. AF was supported by a National Health \& Medical Research Council CJ Martin
Fellowship (ID: 454797). KDE is a recipient of the Betty Behrens
Research Fellowship at Clare Hall, University of Cambridge. The
experiment was sponsored by GlaxoSmithKline and conducted at the GSK
Clinical Unit Cambridge.

\newpage

\bibliographystyle{apalike}

\end{document}